\documentclass[11pt]{article}
\usepackage{graphicx}
\usepackage{amsmath}     
\usepackage{amssymb}    
\usepackage{epsfig}  
\usepackage{setspace} 
\def \BE {\begin{equation}}
\def \EE {\end{equation}}
\def \BEA {\begin{eqnarray}}
\def \EEA {\end{eqnarray}}

\begin{document}
\title{Freely decaying turbulence and
Bose-Einstein condensation in Gross-Pitaevski model}
\author{Sergey Nazarenko$^*$ and  Miguel Onorato$^\dagger$\\
\\
$^*$ Mathematics Institute,
The University of Warwick, 
\\
Coventry, CV4-7AL, UK
\\
$^\dagger$Dipartimento di Fisica Generale,
 Universit\`a di Torino,\\
  Via P. Giuria, 1 - 10125 -
\\
Torino, Italy
}

\maketitle

\abstract{We study turbulence
and Bose-Einstein condensation (BEC) within the two-dimensional
Gross-Pitaevski (GP) model. In the present work, we compute
decaying GP turbulence  in order to establish
whether BEC can occur without forcing and if there is an intensity
threshold for this process.
We use the wavenumber-frequency plots which
allow us to clearly separate the condensate and the wave components and,therefore, to conclude if BEC is present.
We observe that BEC in such a system happens even for very weakly
nonlinear initial conditions without any visible threshold.
 BEC arises via a growing phase coherence due
to anihilation of phase defects/vortices. We study this process
by tracking of propagating vortex pairs. The pairs loose momentum by
scattering the background sound, which results in gradual decrease of the
distance between the vortices.
Occasionally, vortex pairs
collide with
a third vortex thereby emitting sound, which can
lead to more sudden shrinking of the pairs. After the vortex anihilation
the pulse  propagates further as a dark soliton, and it eventually bursts
creating a shock.} 

\section{Background and motivation}
For dilute gases with large energy occupation numbers 
the Bose-Einstein condensation (BEC) \cite{Anderson,Bradley,Davis}
 can be described by the
Gross-Pitaevsky (GP) equation \cite{Gross,Pitaevsky1961}:
\begin{equation}
 i \Psi_t + \Delta \Psi - |\Psi|^2 \Psi =  0,
\label{gp}
\end{equation}
where $\Psi$ is the condensate ``wave function''.
 GP equation also describes light
behavior in media with Kerr nonlinearities.

Many interesting features were found in GP turbulence
in both the nonlinear optics and BEC contexts
 \cite{SV,ZMR85,tkachev,DNPZ,BS,pomeau}.
Initial fields, if weak, behave as wave turbulence (WT)
where
the main nonlinear process is a  four-wave resonant interaction
described by a four-wave kinetic equation \cite{ZMR85}.
This closure was used in \cite{SV,tkachev,DNPZ} to describe the
initial stage of BEC.
It was also theoretically predicted that the four-wave WT closure will eventually
fail due to emergence of a coherent condensate state which is uniform
in space \cite{DNPZ}.
 At this stage the 
nonlinear dynamics can be represented as interactions of small perturbations
about the condensate state. Once again, one can use WT to describe such a system,
but now the leading process will be a three-wave interaction of acoustic-like
waves on the condensate background \cite{DNPZ}. 
Coupling of such acoustic turbulence to the condensate was considered in
\cite{zakhnaz} which allowed to derived the asymptotic law of the condensate growth.

In \cite{NAZ06}, the stage of transition from the
four-wave to the three-wave WT regimes, which itself is a strongly nonlinear
process involving a gas of strongly nonlinear vortices,
 was studied. These vortices anihilate
and their number reduces to zero in a finite time, marking a finite-time 
growth of the correlation length of the phase of $\Psi$ to infinity.
This is similar
to the Kibble-Zurek mechanism of the early Universe phase transitions
which has been introduced originally in cosmology \cite{kibble,zurek}.
It has been established that the vortex anihilation process is aided by the presence
of sound and it becomes incomplete if sound is dissipated.
Fourier transforms in both space and time were analysed
using the wavenumber-frequency plots which, in case of weak wave turbulence, are
narrowly concentrated around the linear dispersion relation $\omega = \omega_k$.
At the initial stage, narrow $(k, \omega)$-distributions around $\omega = \omega_k=k^2$, were
seen
whereas at late evolution stages we saw two narrow components: a
condensate at horizontal line $\omega = \langle \rho \rangle
 = \langle |\Psi|^2 \rangle$ 
and an acoustic component in proximity of the Bogolyubov curve $\Omega_k=  
\langle \rho \rangle + \sqrt{k^4+2\langle \rho \rangle k^2}$.

In \cite{NAZ06},  the system was continuously forced
at either large
or small scales because this is a classical WT setting. 
 WT predictions were confirmed for the energy spectra of GP turbulence.
However, it remained unclear if presence of forcing is essential for complete
BEC process, and whether there is any intensity threshold for this process
in the absence of forcing. This questions are nontrivial because, in principle,
even a weakly forced system could behave very differently from the forced one 
due to an infinite supply of particles over long time.

In the present paper we will examine these questions via numerical simulations
of the 2D GP model without forcing. In addition, we will carefully examine and describe
 the essential stages of the typical route to the vortex anihilation
leading to BEC. In many ways our work is closely related with
numerical studies of decaying 3D GP turbulence of Ref. \cite{BS},
where transition from the 4-wave WT regime to condensation and superfluidity
was studied by visualising emergence and decay of a superfluid vortex tangle.
However, our work addresses additional issues which were not discussed 
in the previous papers, particularly emergence and dispersive properties
of the 3-wave acoustic turbulence, crucial role of sound for the vortex
decay process, separating the wave and the condensate components using 
novel numerical diagnostics based on $(k, \omega)$-distributions.

Below, we will only describe our numerical setup and results. 
For a summary of WT theory and its predictions in the GP context
 we reffer to our previous paper \cite{NAZ06}.

\section{Setup for numerical experiments}

In this paper we consider a setup corresponding to homogeneous
turbulence and, therefore, we ignore finite-size effects due
to magnetic trapping in BEC or to the finite beam radii in
optical experiments.
 For numerical simulations, we have used a standard  pseudo-spectral
method  \cite{FORN} for the 2D equation (\ref{gp}): the nonlinear term is  computed in physical space while  the linear part
is solved exactly in Fourier space. The integration in time
is performed using a second-order Runge-Kutta method.
 The number of grid
points in physical space was set to $N \times N$ with $N=256$. 
Resolution in Fourier space was $\Delta k=2\pi/N$.
Sink  at high wave numbers was provided by adding 
to the right hand side of  equation (\ref{gp}) 
the hyper-viscosity term  $\nu(-\nabla^2)^n \psi$.
Values of $\nu$ and $n$ were selected in order
to localized as much as possible
dissipation to high wave numbers but avoiding 
at the same time the bottleneck effect, - a numerical
artifact of  spectrum pileup at the smallest scales 
\cite{falkovich}.
Note that importance of introducing the small-scale
dissipation to eliminate the bottleneck effect
has long been realised in numerical simulations of
classical Navier-Stokes fluids, and it was also recently
realised in the context of GP turbulence in \cite{KT}.  
We have found, after a number of trials,
that $\nu=2 \times 10^{-6}$ and $n=8$ were
good choices for our purposes. 
Time step for integration was depended on the initial conditions.
For strong nonlinearity smaller time step were required.
 Numerical simulations were performed on a PowerPC G5, 2.7 Ghz.
Initial conditions were provided by the following:
\begin{equation}
\psi(k_x,k_y,t=0)=\frac{\alpha}
 {\sqrt{\pi^{1/2} \sigma}}
  e^{-\frac{(k-k_0)^2} {2\sigma^2}}
 e^{i \phi}
\end{equation}
where $k =\sqrt{k_x^2+k_y^2}$.  
$\alpha$
is a real number which was varied in 
order to change the nonlinearity of the initial condition for different simulations.
$\phi=\phi (k_x,k_y) $
are uniformly distributed random numbers in the interval
$[0,2 \pi]$.
The simulations that will be presented here have been obtained with 
$k_0$= 35 $\Delta k$
and $\sigma=5\Delta k$. 
The nonlinearity of the initial condition 
was measured as $\varepsilon$=$k_0 \alpha$. We have performed simulations ranging from $\varepsilon_{min}=0.018$ to 
$\varepsilon_{max}=1.3$, so we have spanned almost 2 decades of nonlinear parameter in the initial conditions.

\section{Numerical results}
\subsection{Evolving spectra} 

\begin{figure}
\centerline{\includegraphics[width=0.5\textwidth]{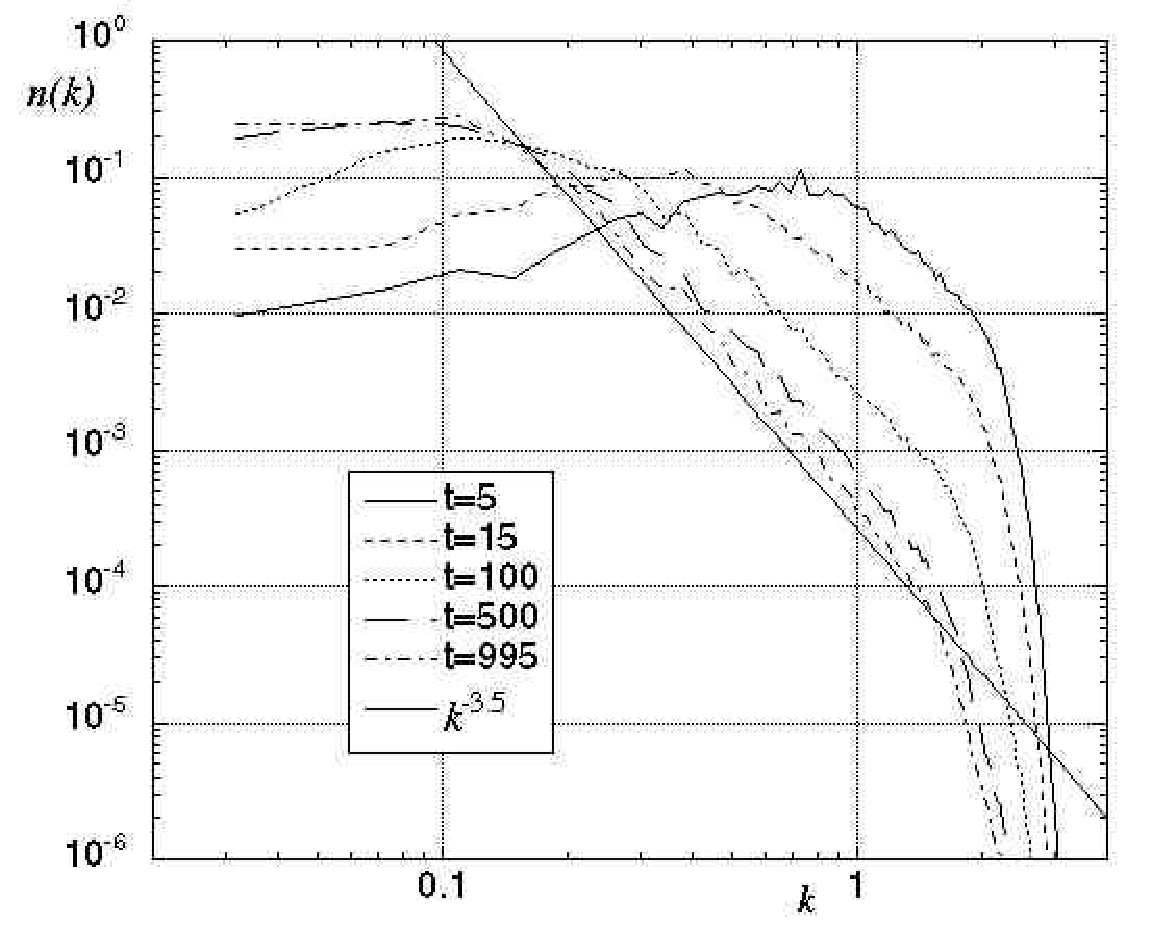}}
\caption{Spectra at different times for the initial condition characterized by 
$\varepsilon=1.3$}\label{fig:spettria}
\end{figure}

\begin{figure}
\centerline{\includegraphics[width=0.5\textwidth]{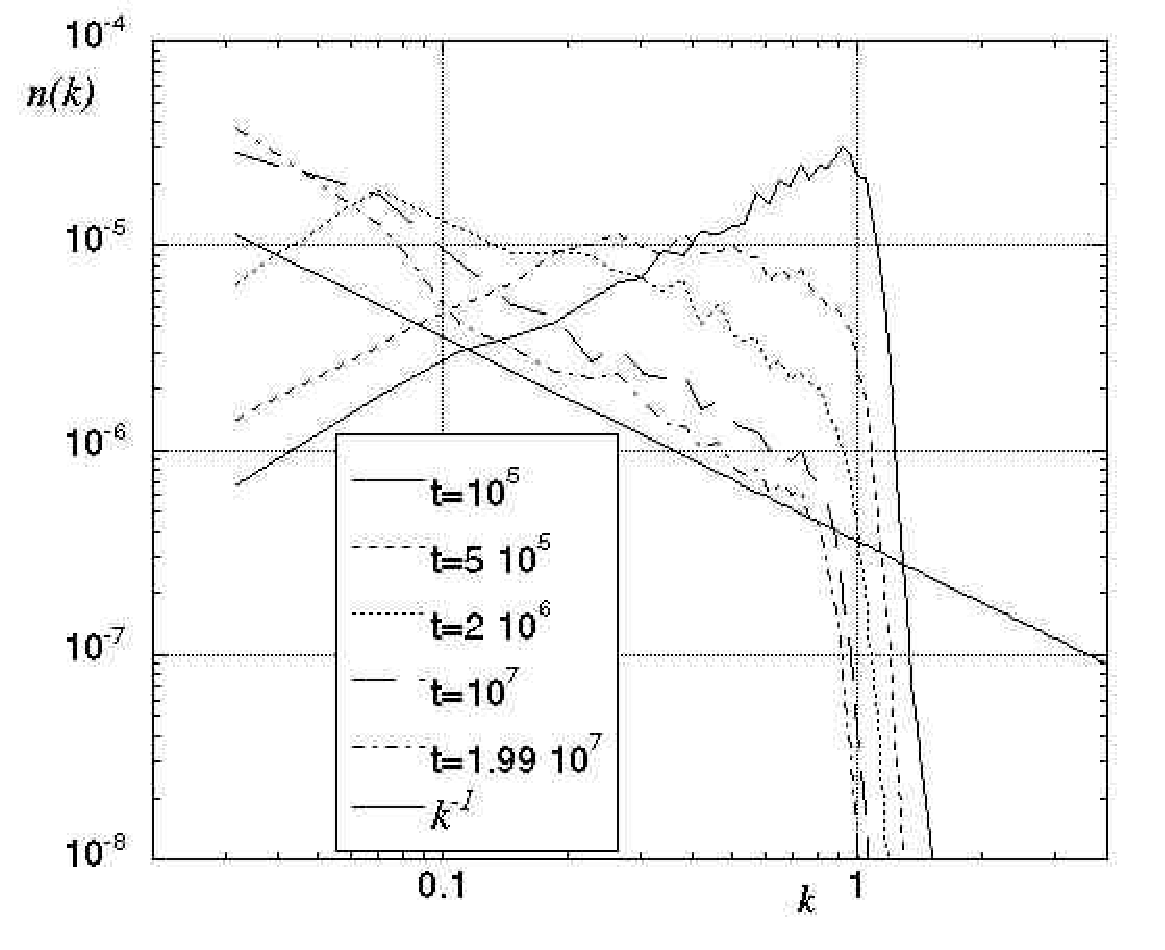}}
\caption{Spectra at different times for the initial condition characterized by 
$\varepsilon=0.018$}\label{fig:spettrib}
\end{figure}

We start by examining the most popular turbulence
object, the spectrum, $$n_k = \langle |\Psi_k |^2 \rangle. $$
 Since we study the setup
corresponding to BEC, we start with a spectrum concentrated
at high wavenumbers leaving a range of smaller
wavenumbers initially empty so that it could be filled
during the evolution.
In figures \ref{fig:spettria} and  \ref{fig:spettrib} 
 we show the spectra at different times 
  for the strongest and the weakest nonlinear initial condition
we have analyzed.

At early stages, for both small and large initial intensities,
we see propagation of the spectrum toward lower wavenumbers.
However, we do not 
observe formation of a  scaling range
corresponding to the wave energy equipartition
$n_k \sim 1/\omega_k = 1/k^2$ (i.e. energy density
$\omega_k n_k$ is constant in the 2D $k$-space)
as it was the case in the simulations with continuous forcing
\cite{NAZ06}.

At later stages, no matter how small the initial
intensity is, the low-k front reaches the smallest
wavenumber and we observe steepening reaching slope
$\sim -3.5 $ for large initial intensities (figure \ref{fig:spettria})
and $\sim -1 $ for the weakest initial data (figure \ref{fig:spettrib}).
This corresponds to WT breakdown and
onset of BEC. However, the information contained in 
spectrum $n_k$ is very incomplete as it does not allow
to distinguish between the coherent condensate and
random waves that may both occupy the same wavenumber
range. Thus, we turn to study the direct measures of condensation
such as the correlation length and the wavenumber-frequency
plots.

\subsection{Explosive growth of correlation length} 

\begin{figure}
\centerline{\includegraphics[width=0.4\textwidth]{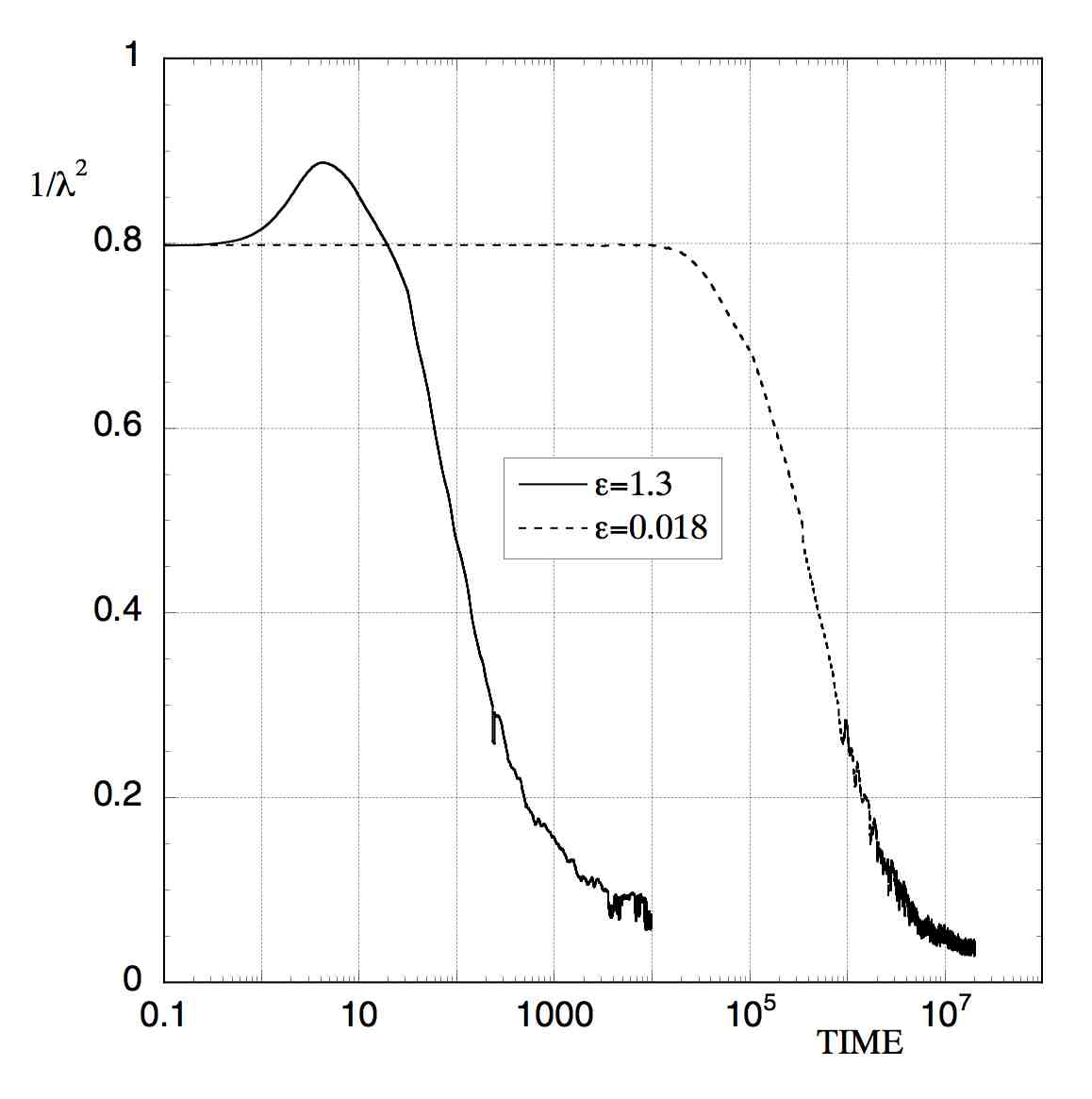}}
\caption{Evolution in time
of the correlation length for different nonlinearity.}\label{fig:log-lin-correlation}
\end{figure}

By definition, condensate is a coherent structure whose correlation
length is of the same order as the bounding box.
We define the correlation length
directly based on the auto-correlation function of field $\Psi$,
\begin{equation}
C_{\psi}(r) = \langle {\cal R} \Psi({\bf x}) \, {\cal R}  \Psi({\bf x + r}) \rangle /
\langle {\cal R} \Psi({\bf x})^2\rangle,
\end{equation}
where ${\cal R} \Psi$ denotes the real part of $\Psi$ (result based on the
imaginary part would be equivalent).
Correlation length $\lambda$ can be defined as
\begin{equation}
\lambda^2 = \int_0^{r_0} C_{\psi}(r) \, d{\bf r},
\end{equation}
where $r_0$ is the first zero of $C_{\psi}(r)$.
Note that initially $C_{\psi}(r)$ can strongly oscillate,
which is a signature of weakly nonlinear
waves.  However, only one oscillation (i.e. within the first
zero crossing)
is relevant to the condensate, which explains our definition of $\lambda$. 
Figure \ref{fig:log-lin-correlation} 
 shows evolution of $1/\lambda^2$ which, as we see
that $\lambda$ always reaches the box size
wich is a signature of BEC.

\subsection{Wavenumber-Frequency plots} 

As we discussed above, the spectra cannot distinguish between
random waves and coherent structures especially when they
are present simultaneously and overlap in the $k$-space.
Besides, the spectra do not tell us if the wave component 
is weakly or strongly nonlinear.
To resolve these ambiguities, following \cite{NAZ06}, let us
perform an additional Fourier transform over a window
of time and examine the resulting 
$(k, \omega)$-plots of space-time Fourier coefficients.


\begin{figure}
\centerline{\includegraphics[width=0.4\textwidth]{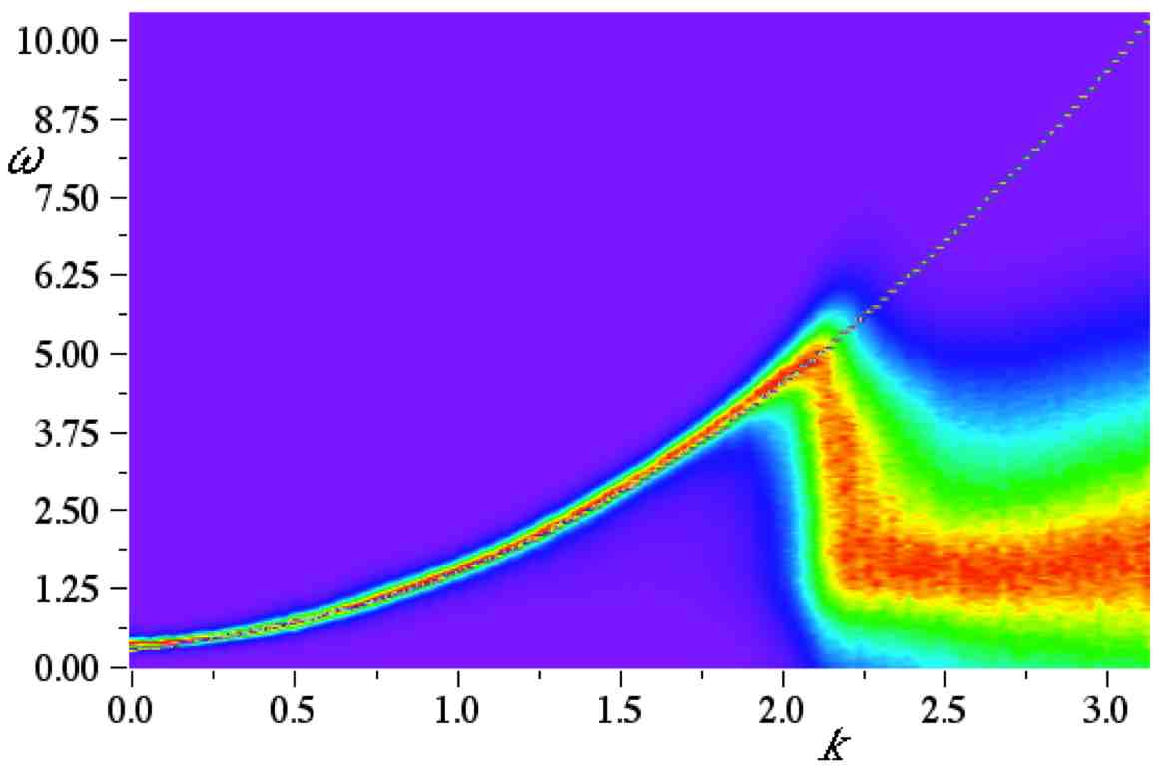}}
\caption{($k, \omega)$ plot for the initial  stage
for the case of  $\varepsilon=0.4$.
 Solid curve shows the Bogolyubov dispersion relation.}
\label{fig:bog_weak_initial}
\end{figure}
\begin{figure}
\centerline{\includegraphics[width=0.4\textwidth]{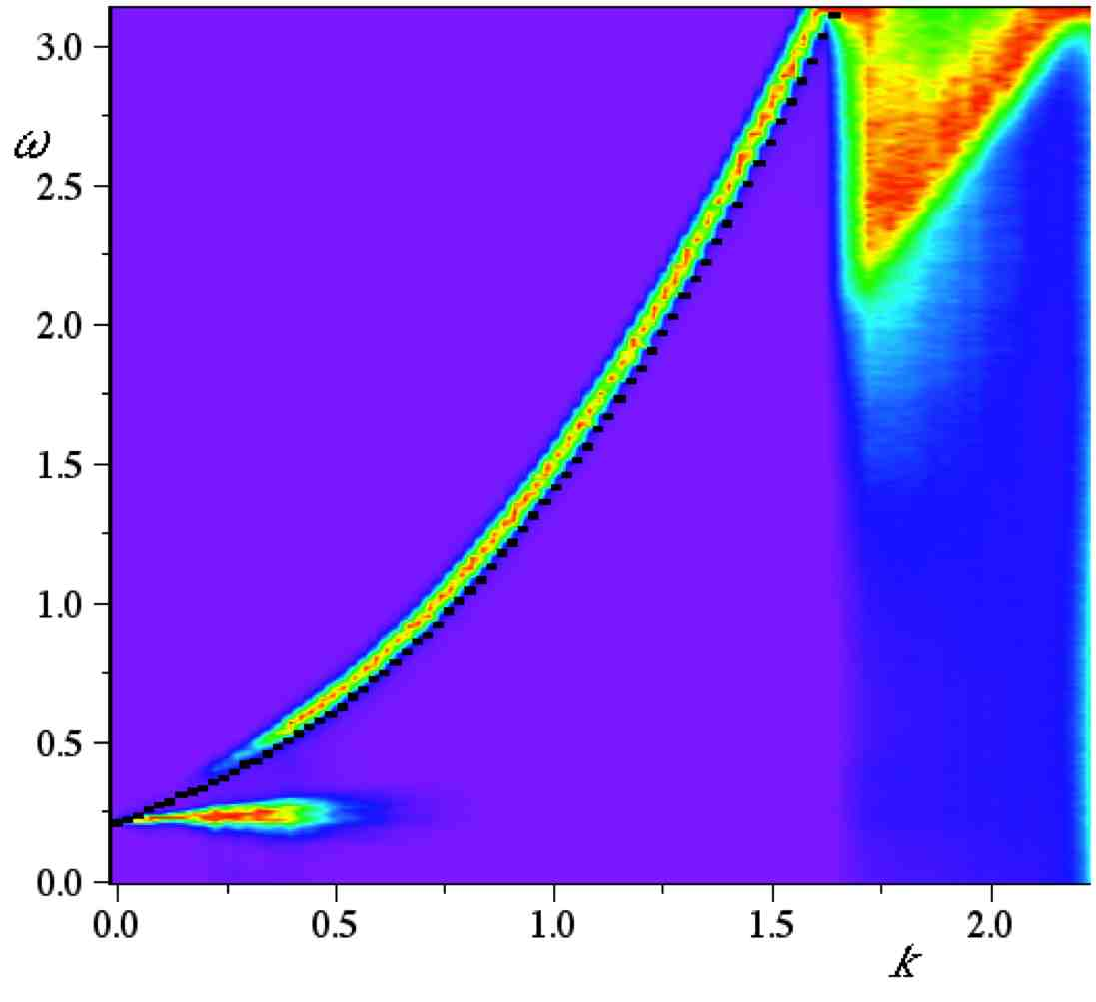}}
\caption{($k,\omega)$ plot for the late  stage
for the case of  $\varepsilon=0.4$.  
Solid curve shows the Bogolyubov dispersion relation.}
\label{fig:bog_weak_final}
\end{figure}

Figure \ref{fig:bog_weak_initial} corresponds to an early time of the system with
relatively weak initial intensity ($\varepsilon =0.4$).
We see that the distribution is narrowly concentrated near
$\omega = k^2$ which indicates that these waves are weakly
nonlinear. The weak nonlinear effects manifestate themselves
in a small up-shift and broadening of the $(k, \omega)$-distribution
with respect to the $\omega = k^2$ curve.
For sufficiently small initial intensities,
these early stages of evolution are characterised
by weak 4-wave turbulence. The breakdown 
of the $\omega = k^2$ curve at high $k$'s
occurs due to the the numerical dissipation
 in the region close to the maximal
wavenumber (this component  is weak but  clearly visible 
 because the color map is normalised to the maximal value 
of the spectrum at each fixed $k$).

\begin{figure}
\centerline{\includegraphics[width=0.4\textwidth]{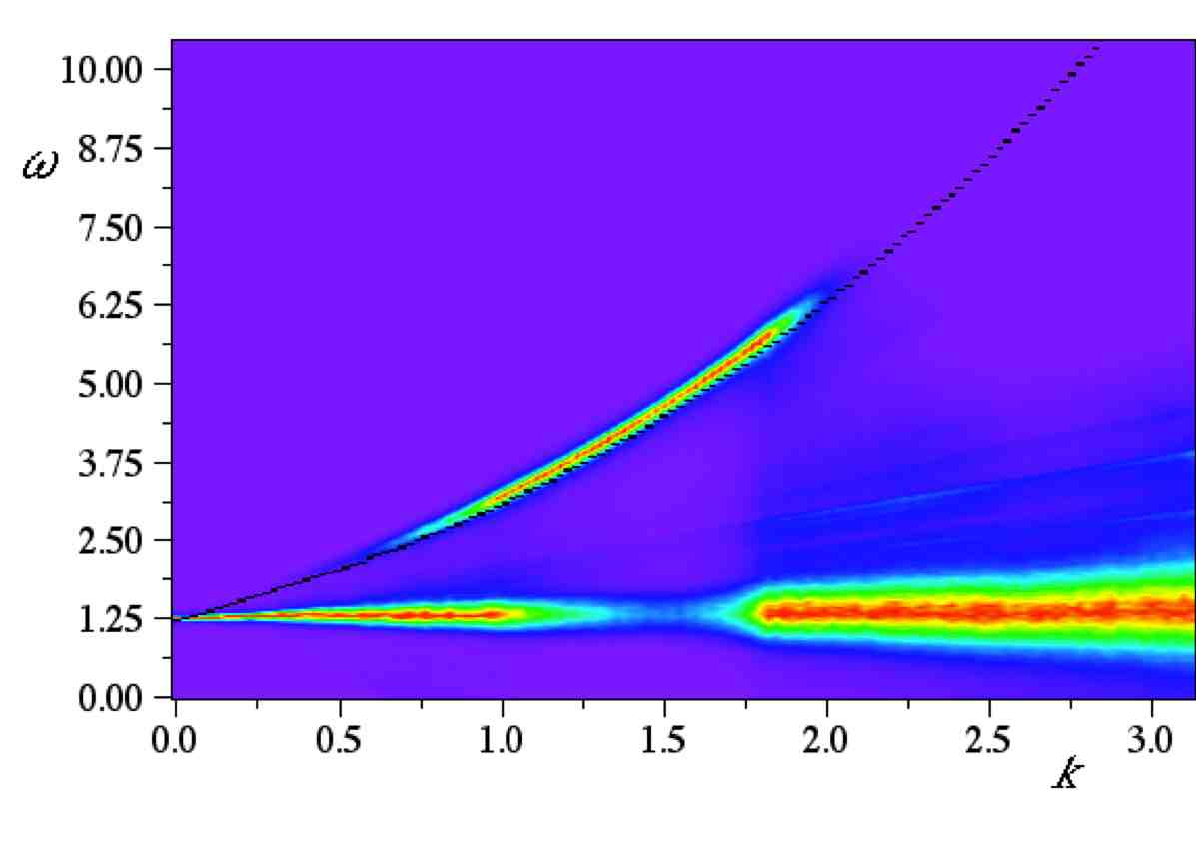}}
\caption{($\omega-k)$ plot for the latest stages
for the most nonlinear case, $\varepsilon=1.3$. 
 Solid curve shows the Bogolyubov dispersion relation.}
\label{fig:bog_strong}
\end{figure}

Figure  \ref{fig:bog_weak_final} shows a late-time plot for the same run (i.e. $\varepsilon =0.4$).
The late-stage $(k, \omega)$-plots for the most nonlinear 
intensities are in shown in figure \ref{fig:bog_strong}.
We see that in both cases we now see two clearly separated components
quite narrowly concentrated around the following curves:
\begin{itemize}
\item
(A) A horizontal line with $\omega \approx \langle \rho \rangle$,
\item
(B) The upper curve which follows the Bogolyubov curve
$\omega = 
 \Omega_k=       
\langle \rho \rangle + \sqrt{k^4+2\langle \rho \rangle k^2}$.
\end{itemize}
Component (A) corresponds to BEC. Its coherency can be seen
in the fact that the frequency of different wavenumbers
is the same. Note that usually BEC is depicted as a component
with the lowest possible wavenumber in the system, whereas in
our case we see a spread over, although small, but finite
range of wavenumbers. This wavenumber spread is caused by
few remaining deffects/vortices.

Curve (B) corresponds the Bogolyubov sound-like waves.
We see that the the $(k, \omega)$-distribution
is quite narrow and close to the Bogolyubov dispersion curve,
 which indicates
that these waves are weakly nonlinear. However, now these 
weakly nonlinear waves travel on a strongly nonlinear
BEC background. This is a three-wave acoustic weak
turbulence regime \cite{DNPZ, zakhnaz, NAZ06}.

\subsection{Separating the condensate and the wave components}

\begin{figure}
\centerline{\includegraphics[width=0.6\textwidth]{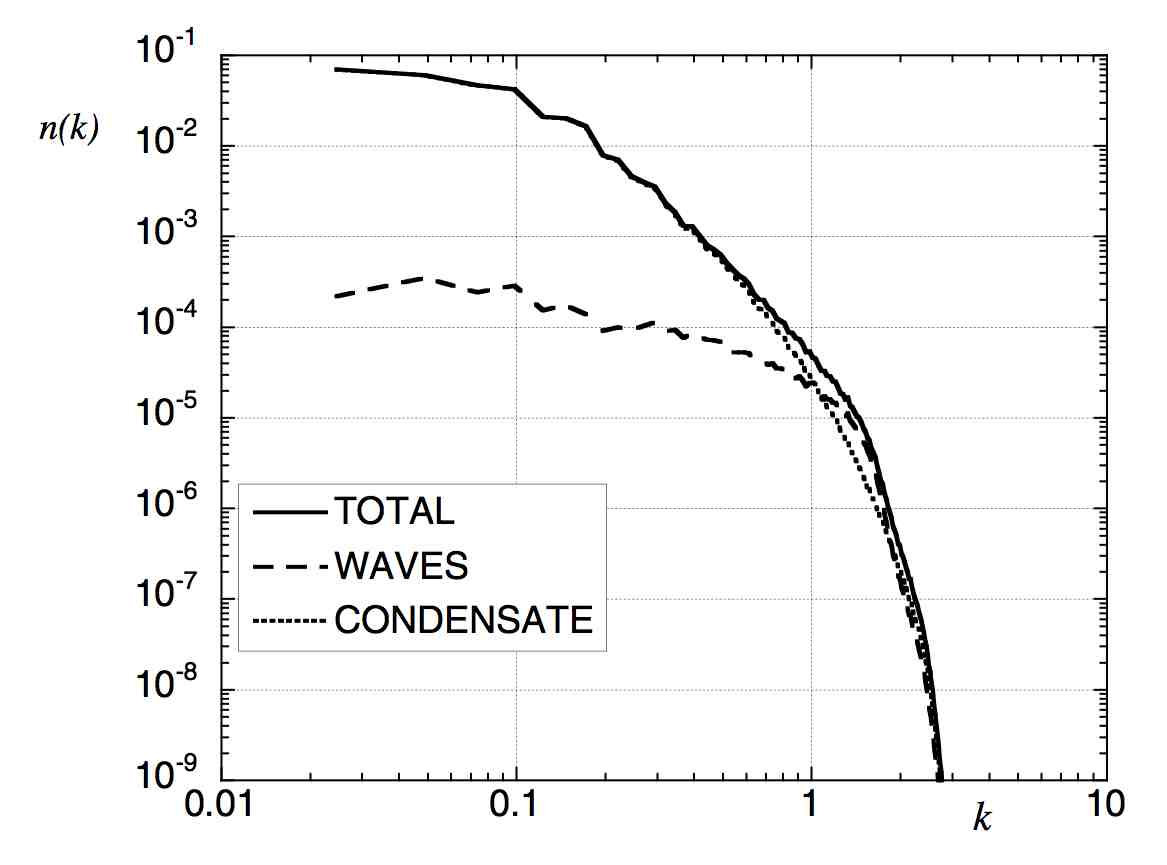}}
\caption{Total wave spectrum (solid line), 
condensate wave spectrum (dotted line) and
background wave spectrum (dotted line)  }
\label{fig:filter_spectra} 
\end{figure}

Using the Wavenumber-Frequency plots we can
separate BEC and the wave component
and plot their spectra separately. 
In figure \ref{fig:filter_spectra} we show the spectrum
for the case of $\varepsilon=1.3$ at late time for the condensate 
and the rest of the wave field. These spectra have been obtained by 
integrating the $(\omega-k)$ from 0 to a threshold $\omega_c$
(this corresponds to condensate) and from $\omega_c$ to 
the maximum value of $\omega$ considered. In the 
present case, in order to separate the condensate, $\omega_c$
was set to 1.5 (see  figure \ref{fig:bog_strong}).
As is clear from the the figure, most of the energy is concentrated in
the condensate.

\subsection{Typical events leading to vortex anihilation}

\begin{figure}
\centerline{\includegraphics[width=0.8\textwidth]{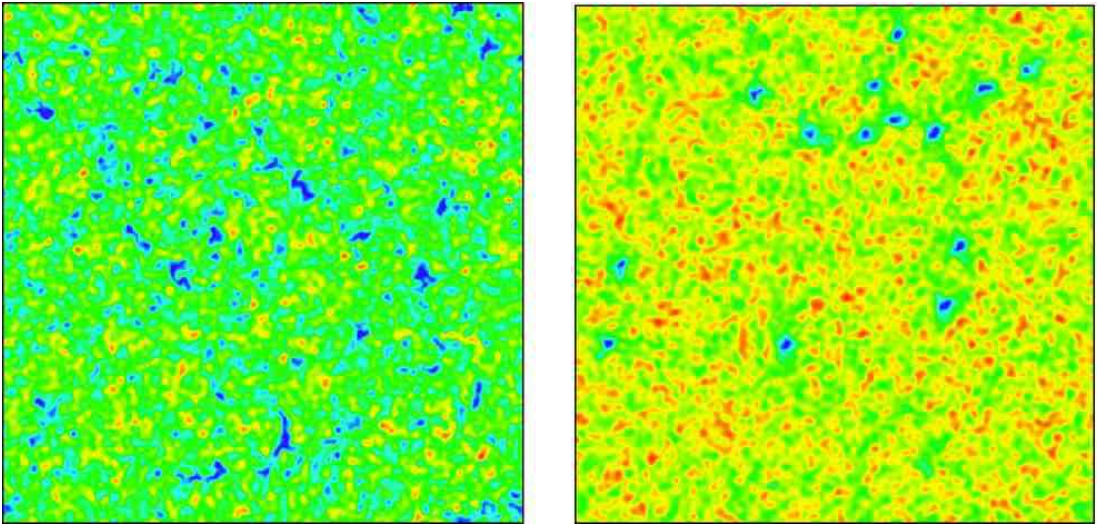}}
\caption{$|\Psi(x,y,t))|$ at two different instants of time}
\label{fig:vortices} 
\end{figure}

Vortices ar phase deffects of $\Psi$ and, therefore,
the correlation length growth is intimately connected
with the decrease of the total number of vortices.
As argued  in \cite{NAZ06}
that $N_{vortices} \sim 1/\lambda^2$.
To give an illustration of the anihilation process we show in figure 
\ref{fig:vortices} two snapshots of $|\Psi(x,y,t)|$ at different times.
The vortices are seen as blue spots in these snapshots and we see a considerable
decrease of their number at the later time (on the right).
It was shown in \cite{NAZ06} that Bogolyubov sound is 
an essential mediator in the vortex anihilation process,
and that introducing a sound absorption can lead to
frustration and incompleteness of the BEC process.

\begin{figure}
\centerline{\includegraphics[width=0.5\textwidth]{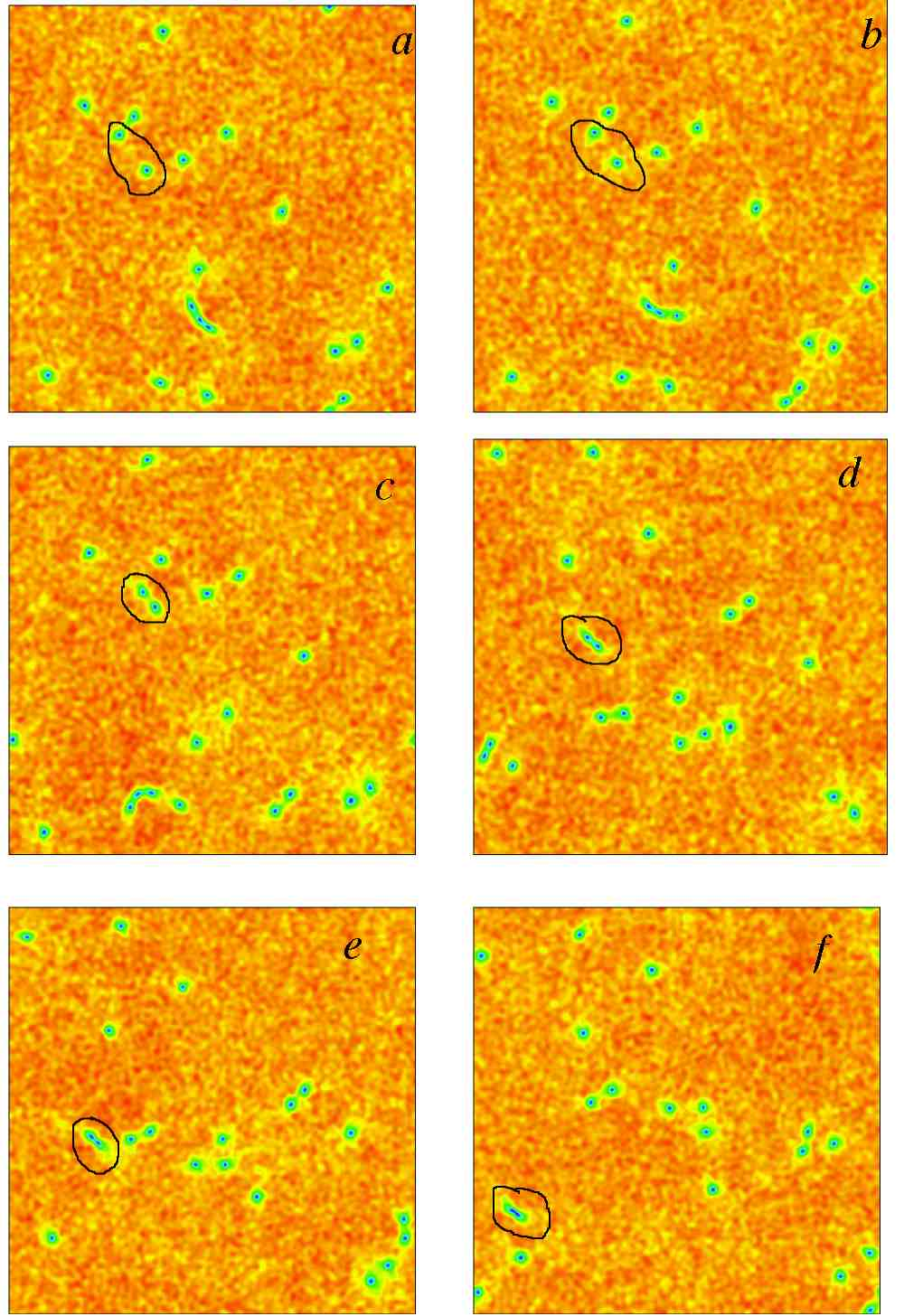}}
\caption{}
\label{fig:movie1} 
\end{figure}

\begin{figure}
\centerline{\includegraphics[width=0.5\textwidth]{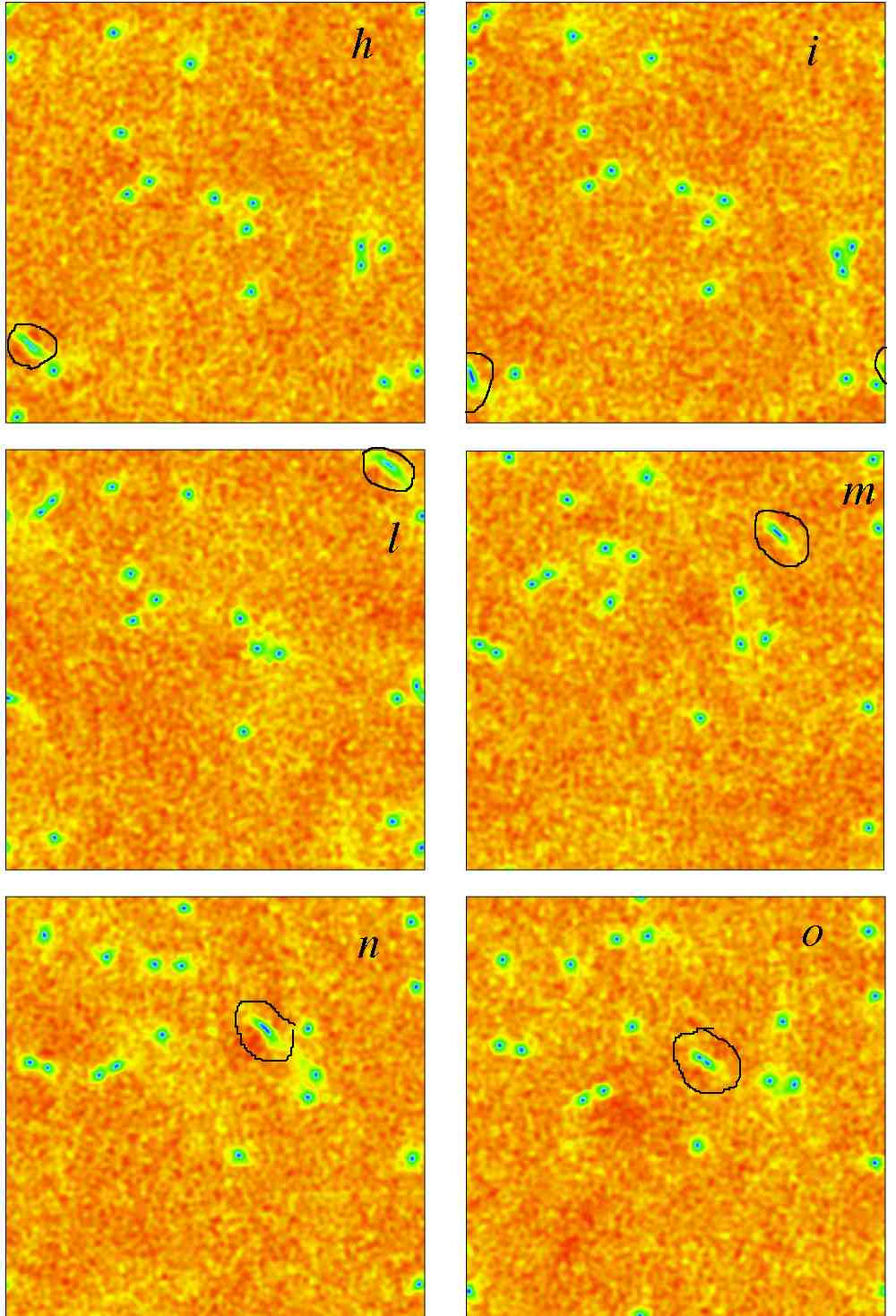}}
\caption{}
\label{fig:movie2} 
\end{figure}

\begin{figure}
\centerline{\includegraphics[width=0.5\textwidth]{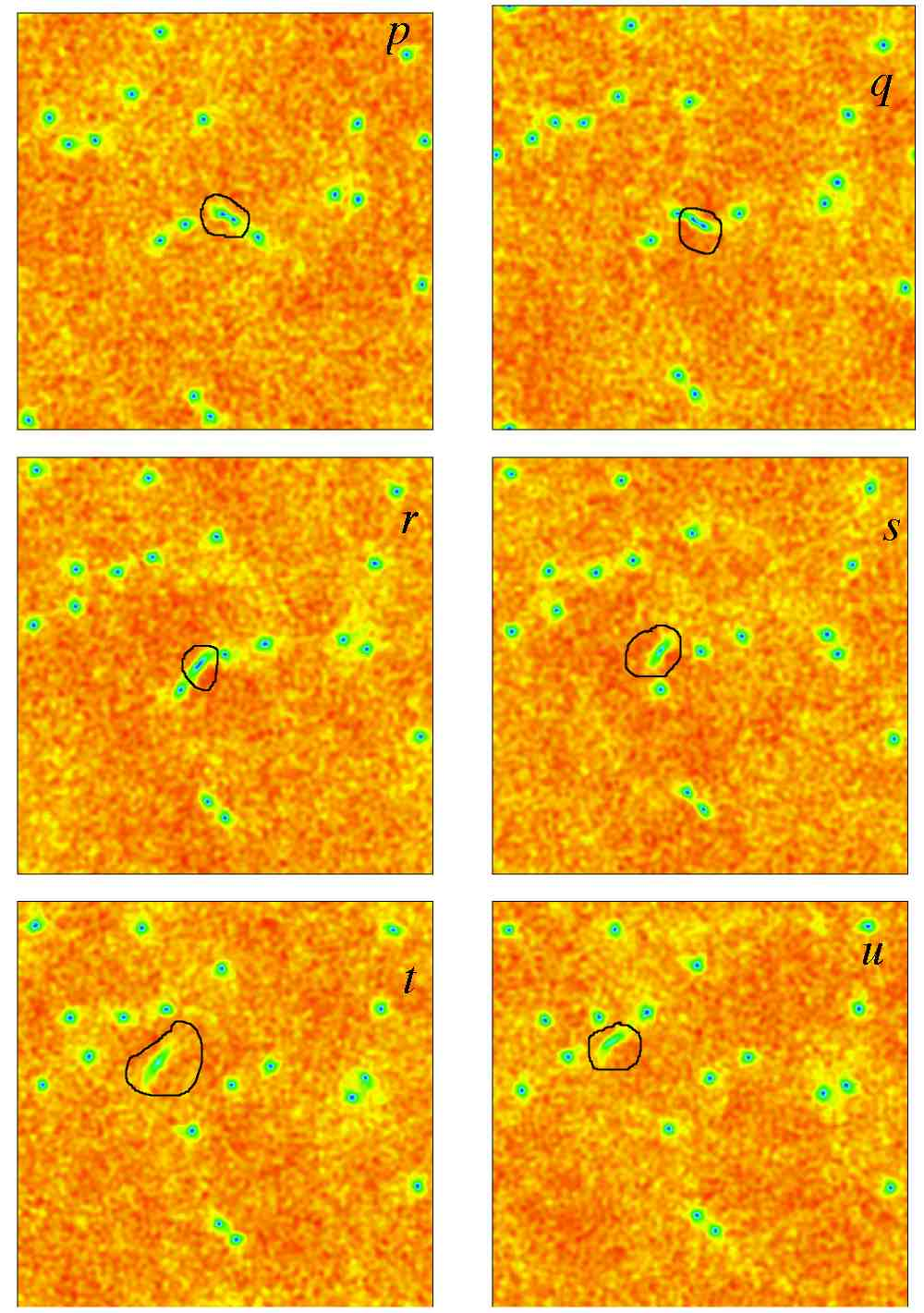}}
\caption{}
\label{fig:movie3} 
\end{figure}

Let us now examine in detail the typical sequence of events
leading to the vortex annihilation by tracking vortex
pairs that are destined to anihilate. 
The vortex pair motion is best seen in a computer generated
movie which is available upon request from the authors.
In Figures \ref{fig:movie1},  \ref{fig:movie2} and \ref{fig:movie3}
we show a representative sequence of frames from this 
movie.

\begin{figure}
\centerline{\includegraphics[width=0.5\textwidth]{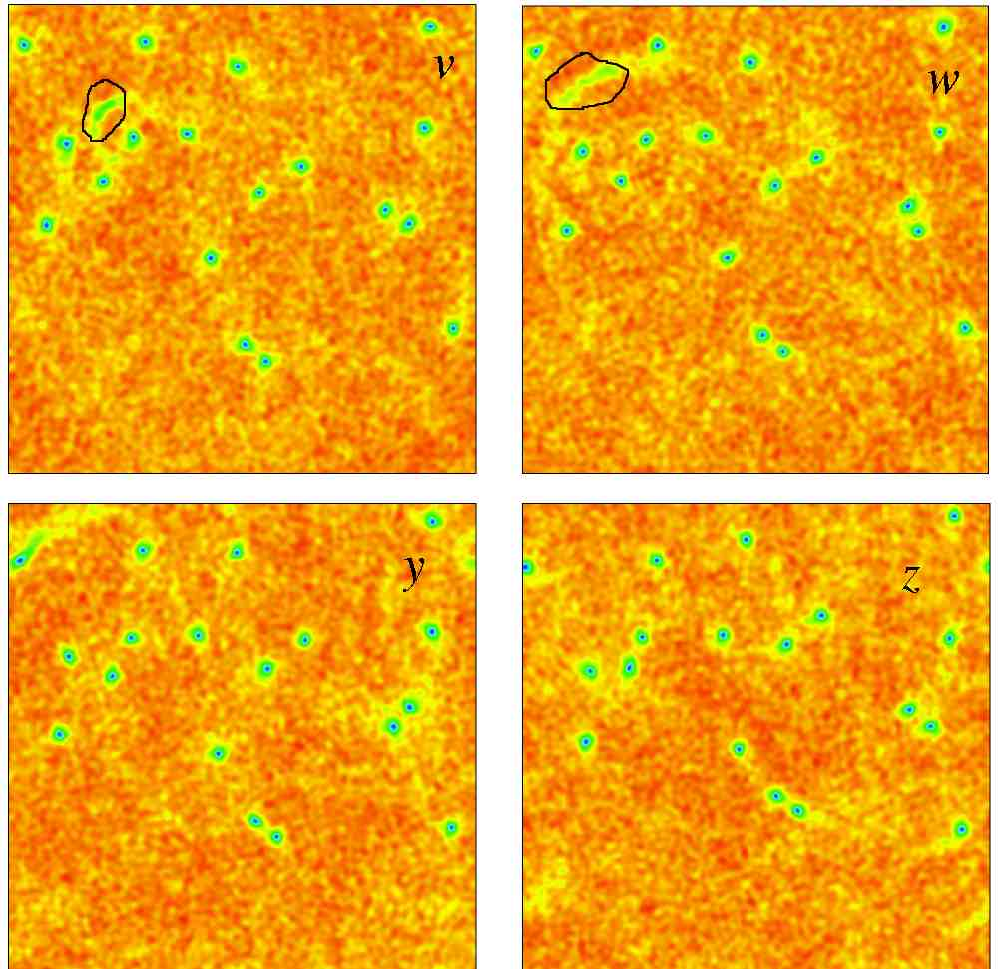}}
\caption{}
\label{fig:movie4} 
\end{figure}
We can see that the vortex pair forms in frames a to c
so that the distance between the vortices in the
pair is considerably less than distance to the
other vortices. Such a pair propagates  
like a vortex dipole in fluid, as seen in frames c to f.
During this propagation, the vortex pair scatters
the ambient sound waves thereby transferring its
momentum to the acoustic field. This momentum loss
make vortices get closer to each other and, therefore,
move faster as a pair. This process can be interpreted
as a ``friction'' between the vortices and a
``normal component'' (phonons). It is easy to show
that such process leads to the change of distance $d(t)$
between the vortices like 
\begin{equation}
d(t) = \alpha \sqrt{t^A -t},
\label{d}
\end{equation}
where coefficient $\alpha$ is proportional to the energy
density of sound and $t^A$ is the anihilation time.
\begin{figure}
\centerline{\includegraphics[width=0.5\textwidth]{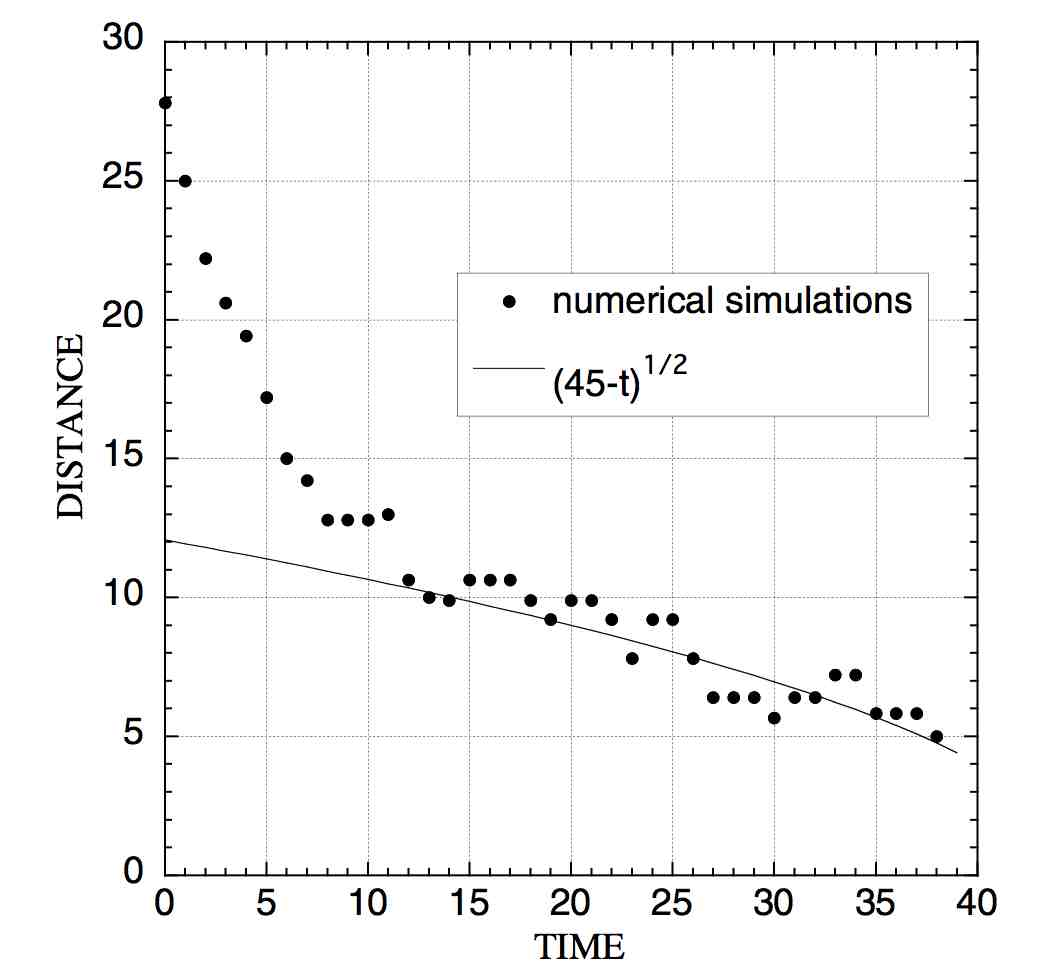}}
\caption{Distance between a pair vortices as a function 
of time for the vortex pair and
the time range of Figure \ref{fig:movie1}.}
\label{fig:distance} 
\end{figure}
Figure \ref{fig:distance}
 shows evolution of the inter-vortex distance $d(t)$,
 calculated 
for the vortex pair and
the time range of Figure \ref{fig:movie1}.
We see that for the time range when the vortex pair
is more or less isolated from the other vortices, $11<t<40$, the
inter-vortex distance shrinks in qualitative agreement with
law (\ref{d}). However, immediately after that, in frame h in
Figure \ref{fig:movie2}, the vortex pair collide with a third vortex
and suddenly shrink and anihilate.
Further on the movie we see that 
 the vortex pair momentum was not completely
lost and it keep propagating a Jones-Roberts dark
soliton \cite{jones_roberts}.
In between of frames h and o the soliton ``cannot make up
his mind'' oscillating between the state with and without vortices
near the vortex anihilation threshold, until frame p where it re-emerges
as a vortex pair. It collides with yet another vortex and changes
its propagation direction to 90 degree in between
of frames q and r, it annihilates again in frame s, propagates
as a dark soliton until frame v.
Eventually, this soliton is weakened due
to further sound generation and scattering and 
it becomes too weak to maintain its stability and integrity.
At this point it bursts thereby generating a shock
wave as seen in frames v to y.

Summarizing,  we can identify the following
important events on the route to vortex anihilation:
\begin{itemize}
\item
Gradual shrinking of the inter-vortex distance
due to the sound scattering when the vortex pair is sufficiently
isolated from the rest of the vortices,
\item 
Sudden shrinking events due to collisions with a third
vortex and resulting sound generation (similar three-vortex 
event was described in \cite{bare}),
\item
Post-anihilation propagation of dark solitons,
\item
Occasional recovery of vortex pairs 
in dark solitons which are close to the critical amplitude,
\item
Weakening and loss of stability of the dark solitons resulting in
a shock wave.
\end{itemize}

\section{Conclusions}

We have computed decaying GP turbulence 
in a 2D periodic box with initial spectrum
occupying the small-scale range. We observed that
BEC at large scales arises for any initial intensity
without a visible threshold, even for very weakly nonlinear
initial conditions. BEC was detected by analysing the spectra
and the correlation length and, most clearly, by analysing the
$(k, \omega)$ plots.  On these plots BEC and Bogolyubov waves are
seen as two clearly distinct components: (i) BEC coherently oscillating
at nearly constant frequency and (ii) weakly nonlinear waves closely
following the Bogolyubov dispersion.
By separating BEC and the waves we observed that most of the
energy at late times is residing in the BEC component at most of the 
important scales except for the smallest ones.

We also analysed the typical events on the path to vortex anihilation, -
an essential mechanism of BEC.  

\noindent {\bf Acknowledgments} \\
Al Osborne is acknowledged for discussions in
 the early stages of the work.

\end{document}